\documentclass[journal,draftclsnofoot, perreview,12pt,onecolumn]{IEEEtran}
\usepackage{amsfonts}
\usepackage{mathrsfs}
\usepackage{multirow}

\usepackage{bbm}
\usepackage{amsmath}
\usepackage{multirow}
\usepackage{booktabs}
\usepackage{setspace}
\usepackage{slashbox}
\usepackage{amssymb,amscd,amsmath}
\usepackage[all]{xy}
\usepackage{fancyhdr}
\usepackage{subfigure}
\usepackage{citesort}
\usepackage{threeparttable}
\usepackage{graphicx}

\ifCLASSINFOpdf
\else
\fi

\begin{document}
%
\title{Partially Block Markov Superposition Transmission of Gaussian Source with Nested Lattice Codes}

%
%

\author{{Shancheng Zhao and Xiao Ma}
\thanks{Shancheng Zhao is with the College of Information Science and Technology, Jinan University, Guangzhou, China~(shanchengzhao@jnu.edu.cn).

Xiao Ma is with the School of Data and Computer Science, Sun Yat-sen University, Guangzhou, China~(maxiao@mail.sysu.edu.cn).

This work was partially supported by the NSF of China~(No.~91438101 and No.~61501206) and the 863 Program~(No.~2015AA01A709).
}
}

\maketitle

\begin{abstract}
This paper studies the transmission of Gaussian sources through additive white Gaussian noise~(AWGN) channels in bandwidth expansion regime, i.e.,~the channel bandwidth is greater than the source bandwidth. To mitigate the error propagation phenomenon of conventional digital transmission schemes, we propose in this paper a new capacity-approaching joint source channel coding~(JSCC) scheme based on {\em partially} block Markov superposition transmission~(BMST) of nested lattice codes. In the proposed scheme, first, the Gaussian source sequence is discretized by a lattice-based quantizer, resulting in a sequence of lattice points. Second, these lattice points are encoded by a {\em short} {\em systematic} group code. Third, the coded sequence is partitioned into blocks of equal length and then transmitted in the BMST manner. Main characteristics of the proposed JSCC scheme include:~1)~Entropy coding is not used explicitly. 2)~Only parity-check sequence is superimposed, hence, termed partially BMST~(PBMST). This is different from the original BMST. To show the superior performance of the proposed scheme, we present extensive simulation results which show that the proposed scheme performs within 1.0~dB of the Shannon limits.
Hence, the proposed scheme provides an attractive candidate for transmission of Gaussian sources.


\end{abstract}


%
\IEEEpeerreviewmaketitle

\section{Introduction}\label{introduction}
%
%
%
%

%


Transmission of analog-valued source through additive white Gaussian noise~(AWGN) channels is of both theoretical and practical importance. Based on modulation techniques used, commonly implemented transmission schemes can be categorized into three types: analog, digital, and hybrid analog-digital~(HDA). Digital transmission scheme, guided by Shannon's separation theorem, consists of source encoder and channel encoder at the transmitter. One of the advantages of digital transmission scheme is that, based on sophisticated source encoder and channel encoder, it performs asymptotically close to the theoretical limit, as was proved in~\cite{Shannon1948A,Berger1971Rate}. However, to achieve the optimality, sufficiently large blocklengths are required, which lead to large buffer and long delay. This is unacceptable for most practical applications. Hence, in practice, source encoder is implemented through the concatenation of a vector quantizer~(VQ) and an entropy encoder, and channel encoder is implemented with high-performance finite-length codes, e.g., turbo codes~\cite{Berrou93} and low-density parity-check~(LDPC) codes~\cite{Gallager62}. To improve performances in finite-length regime, various joint source channel coding~(JSCC) schemes have been proposed. The simplest JSCC scheme is based on separated source encoder and channel encoder, but with rate allocations between them~\cite{Stoufs2008Scalable}. In~\cite{Farvardin87Optimal}, the authors presented a JSCC scheme based on joint optimization of the source quantizer and the channel encoder/decoder pair. They also developed necessary conditions for the joint optimality of the source and the channel encoder pair. If suboptimal source encoder is used, extra redundancy exists in the output of the source encoder. These redundancy can be incorporated into channel decoding~\cite{Wang98Error} to improve the performance. Joint design of the source and channel encoder was investigated in~\cite{Goodman83Combined,Daut83Two}. JSCC scheme based on the concatenation of two LDPC codes are proposed and analyzed in~\cite{Fresia2009Optimized}. More on JSCC schemes can be found in~\cite{Hamzaoui2002Rate,Xu2007Distributed,Fresia2010Joint} and references therein.

Digital transmission scheme suffers from the so-called {\em threshold effect} and the {\em leveling-off effect}. The threshold effect refers to the fact that the distortion degrades drastically when the signal-to-noise ratio~(SNR) is below a certain threshold and the leveling-off effect refers to fact that the distortion remains constant even increasing the SNR. Analogy transmission scheme via direct amplitude modulation offers the simplest way to cope with these two issues. As an important example, it was shown in~\cite{Goblick1965Theoretical} that uncoded transmission, hence analog transmission, of Gaussian sources through AWGN channels achieves the optimal power-distortion trade-offs in matched bandwidth regime, i.e., the source bandwidth is equal to the channel bandwidth. However, in unmatched bandwidth regimes, analog transmission schemes typically perform far away from the Shannon limits. To combine the advantages of digital and analog transmission schemes, various hybrid digital-analog~(HDA) joint source channel coding~(JSCC) schemes have been proposed, see~\cite{Mittal2002Hybrid,Skoglund2002Design,Skoglund2006Hybrid} and the references therein. In~\cite{Mittal2002Hybrid}, the authors analyzed the asymptotic performances of serval HDA-JSCC schemes. In~\cite{Skoglund2002Design}, a low-complexity and low-delay HDA-JSCC scheme based on VQ was presented for bandwidth expansion regime, i.e., the channel bandwidth is greater than the source bandwidth. This scheme was extended in~\cite{Skoglund2006Hybrid} for both the bandwidth expansion and the bandwidth compression regime by implementing the digital part with turbo codes. More on analog and HDA schemes can be found in~\cite{Wilson2010Joint,Gao2010New,Hu2011Analog} and references therein.

Another fundamental issue associated with conventional digital transmission schemes is the {\em catastrophic error propagation phenomenon} due to the lack of robustness of entropy coding. That is, a few bit errors after channel decoding may lead to bursty breakdown of the source decoder and then result in severe distortion. This abrupt change of distortion is undesired for some important applications, e.g., deep-space applications and multimedia applications. One simple but inefficient approach to this issue is to add redundant bits intentionally in the entropy coding stage to detect the erroneous bits or to prevent error propagation. In~\cite{Yang2012A}, a different approach was proposed based on the concatenation of VQ and systematic rateless LDPC codes over groups. A scheme similar to that of~\cite{Yang2012A} was proposed in~\cite{Bursalioglu2013Joint} for image transmission in deep-space applications.

In this paper, we attempt to design capacity-approaching digital JSCC scheme for Gaussian sources transmitted through AWGN channels in bandwidth expansion regime. We focus on digital transmission scheme for the following reasons.
\begin{itemize}
  \item First, with advanced quantization and error correction techniques, digital transmission schemes can be designed to perform very close to the Shannon limits.
  \item Second, the leveling-off effect is desired in some applications. For an example, high-quality multimedia applications are required in daily life. In these applications, we expect that the decoded signal quality remains constant even channel conditions are varying, as large variations in the received signal quality over a short periods of time may be annoying to end users.
  \item Third, digital parts of serval high-performance HDA-JSCC schemes~\cite{Skoglund2006Hybrid,Rungeler2014Design} are implemented with digital transmission schemes. Hence, we may improve performances of these HDA-JSCC schemes by designing new high-performance digital transmission scheme.
\end{itemize}
The proposed JSCC scheme is based on partially block Markov superposition transmission~(BMST) of nested lattice codes. In the proposed scheme, first, the Gaussian source sequence is discretized by a lattice-based quantizer, resulting in a sequence of lattice points. Second, these lattice points are encoded by a short systematic group code. Third, the coded sequence is partitioned into blocks of equal length and then transmitted in the BMST manner. Main characteristics of the proposed scheme include:~1)~In the proposed scheme, only the parity lattice points are superimposed, hence, termed partially BMST~(PBMST). This is different from the original BMST. 2)~The proposed scheme avoids the explicit use of entropy coding, hence mitigates the error propagation phenomenon of conventional digital scheme. To assess the performance of the proposed scheme, numerous simulations are presented, which show that the proposed scheme performs close to the Shannon limits~(within 1.0~dB).
Hence, the proposed scheme provides an attractive candidate for digital transmission of analogy source. It may also find applications in some HDA-JSCC schemes proposed in the literature.

\section{Problem Statements}\label{SecII}
Let $\mathbf{S}=(s_0, s_1, s_2,...)$ denote a discrete-time Gaussian source, where $s_i$ is a sample of Gaussian random variable with zero mean and variance $\sigma^2 = 1$. In this paper, we will consider both memoryless Gaussian sources and correlated Gaussian sources. We further assume that the source emits $R_s$ symbols in time interval $T$ and the channel can transmit at most $R_c$ symbols in time interval $T$. If $R_s{\approx}R_c$, which is equivalent to saying that the source and the channel have comparable bandwidth, it has been shown that coding is not required~\cite{Goblick1965Theoretical}. However, if $R_s \ll R_c$, we can always find a coding scheme to achieve a better performance. In this paper, we will assume that $R_s\ll R_c$.

A general coding scheme can be described as follows.

\begin{enumerate}
  \item {\bf Encoding:}
  \begin{equation}
    \phi: (s_0,s_1,...s_{k-1}) \mapsto (x_0,x_1,....x_{n-1}),
  \end{equation}
  where $x_i$'s satisfy the power constraint $\frac{1}{n}\sum_i x_i^2\leq{P}$.
  \item {\bf Transmitting:} Assume that $x_i$ is transmitted and $y_i = x_i + z_i$ is received, where $z_i$ is the additive Gaussian noise.
  \item {\bf Decoding:}
  \begin{equation}
    \psi: (y_0,y_1,...y_{n-1}) \mapsto  (\hat{s}_0,\hat{s}_1,...\hat{s}_{k-1}).
  \end{equation}
\end{enumerate}
The performance of the above~(general) coding scheme can be measured by the average transmitted power $P$ versus the distortion calculated as
 \begin{equation}\label{Distortion}
    D = \frac{1}{k}\sum_i(s_i-\hat{s_i})^2.
 \end{equation}

\begin{figure}[!t]
\centering
\includegraphics[width=12cm]{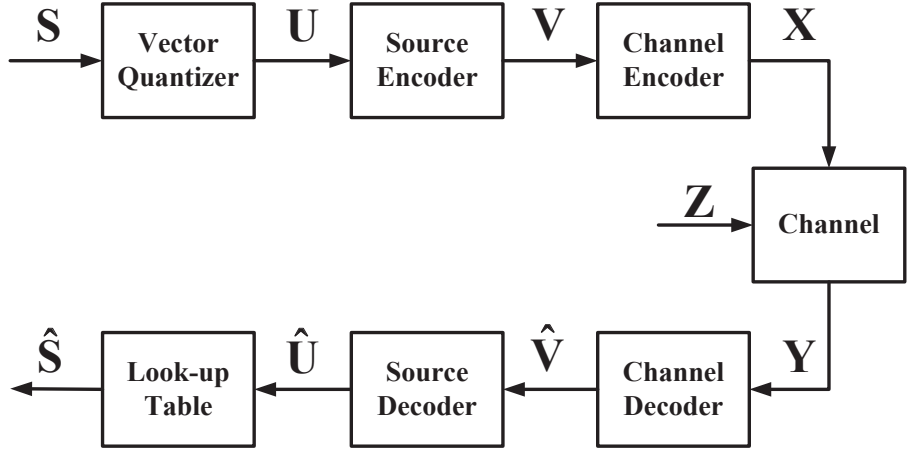}
\caption{The conventional digital transmission scheme.} \label{Conventional}
\end{figure}

Conventionally, the general coding scheme can be implemented as follows, see Fig.~\ref{Conventional} for a reference.

\begin{enumerate}
  \item {\bf Encoding:}
 \begin{equation}
    {\phi}_1: \mathbf{S}\mapsto\mathbf{U},~~{\phi}_2: \mathbf{U}\mapsto\mathbf{V},~~{\phi}_3: \mathbf{V}\mapsto\mathbf{X},
 \end{equation}
 where $\mathbf{U}$, $\mathbf{V}$ and $\mathbf{X}$ are sequences of representative points, binary digits and transmitted signals, respectively.
  \item {\bf Decoding:}
\begin{equation}
    {\psi}_3: \mathbf{Y}\mapsto\mathbf{\hat{V}},~~{\psi}_2: \mathbf{\hat{V}}\mapsto\mathbf{\hat U},~~{\psi}_1: \mathbf{\hat U}\mapsto\mathbf{\hat S},
 \end{equation}
 where $\mathbf{\hat V}$, $\mathbf{\hat U}$ and $\mathbf{\hat S}$ are estimates of $\mathbf{V}$, $\mathbf{U}$ and $\mathbf{S}$, respectively.
\end{enumerate}

Typically, ${\phi}_1$ is implemented with a VQ, ${\phi}_2$ is implemented with an entropy encoding algorithm, and ${\phi}_3$ is implemented a high-performance error correction code. One critical issue associated with this conventional coding scheme is: a few errors in the channel coding stage may cause the breakdown of the source decoder, owing to the lack of robustness of entropy coding. To resolve this issue, we propose in the next section a new JSCC scheme which avoids the explicit use of entropy coding.

\section{The JSCC Scheme Based on Partially Block Markov Superposition Transmission of Nested Lattice Codes}\label{SecIII}
The transmission diagram of the propose JSCC scheme is shown in Fig.~\ref{ProposedJSCC:subfig:a}. First, the Gaussian source sequence $\mathbf{S}$ is quantized with a lattice-based VQ to generate $\mathbf{U}$, a sequence of lattice points. Second, these lattice points are encoded in the partially block Markov superposition transmission~(PBMST) manner~(to be explained later) to obtain the sequence of lattice points $\mathbf{X} = (\mathbf{U}, \mathbf{P})$, which serves as the input to the channel. That is, the sequence $\mathbf{U}$ is directly transmitted through the channel, then followed by parity-check lattice points $\mathbf{P}$.

\subsection{Vector Quantizers Based on Nested Lattice Codes}
{\em Nested Lattice Codes:~} Let $\Lambda \in \mathbb{R}^{\ell}$ be a lattice with $\Lambda'$ as a sub-lattice. A nested lattice code ${\rm G}$ ~\cite{Zamir2002Nested} is defined as the coset code $\Lambda/\Lambda'$ by taking the Voronoi region of $\Lambda'$ as the shaping region. That is, ${\rm G} = \Lambda \bigcap \mathcal{V}(\Lambda')$, where $\mathcal{V}(\Lambda')$ denotes the Voronoi region of $\Lambda'$. We use the notation ${\rm G}$ to emphasize the group structure of nested lattice codes. Note that, to improve the performance of the nested lattice codes, as source code or channel code, a shifted version of the Voronoi region $\mathcal{V}(\Lambda')$ is typically used as the shaping region~\cite{Conway1983A}.

{\em Vector quantizer:~} Without loss of generality, we can rewrite the source as $\mathbf{S} =(s_0,s_1,s_2,...)$, where $s_t \in \mathbb{R}^{\ell}$.
For $s \in \mathbb{R}^{\ell}$, we define a quantizer based on the nested lattice code ${\rm G}$ as
\begin{equation}\label{Quantizer}
    Q(s) = \arg\min\limits_{ v \in {\rm G} }||v - s||^2,
\end{equation}
where $||v - s||^2$ is the squared distance between $v$ and $s$. Evidently, we can define a family of quantizer $Q_{\alpha , \mathbf{v}}$ based on a transformed~(inflation and shifting) nested lattice code, denoted as $\alpha ({\rm G} + \mathbf{v}) = \alpha (\Lambda + \mathbf{v}) \bigcap \mathcal{V}(\alpha (\Lambda'))$, of the original nested lattice codes $G$. In this family, we can choose $\alpha > 0$
and $\mathbf{v}$ such that the distortion $D(\mathbf{S}, \mathbf{V})$ is minimized. The parameters
$\alpha$ and  $\mathbf{v}$ can be optimized to match the characteristics of the source. In this paper, we will present an iterative optimization procedure, similar to the algorithm in~\cite{Vaishampayan1990Optimal}, to determine suboptimal $\alpha$ and $\mathbf{v}$. It should be pointed out that the group structure of nested lattice based quantizer is used in the proposed JSCC scheme. For each $v\in \alpha ({\rm G} + \mathbf{v})$, we define $A_{v}=\{s\in \mathbb{R}^{\ell}:~Q_{\alpha , \mathbf{v}}(s)=v\}$. Hence the probability of $v\in G$ is
\begin{equation}\label{Initial}
    {P}(v)\propto \int_{A_v}f_{S}(s)~{\rm d}s,
\end{equation}
where $f_{S}(s)$ is the ${\ell}$-dimensional joint probability density function of the source.

\subsection{Encoding of Partially Block Markov Superposition Transmission of Nested Lattice Codes}
\begin{figure}
  \centering
  \subfigure[The proposed JSCC scheme.]{
    \label{ProposedJSCC:subfig:a} 
    \includegraphics[width=12cm]{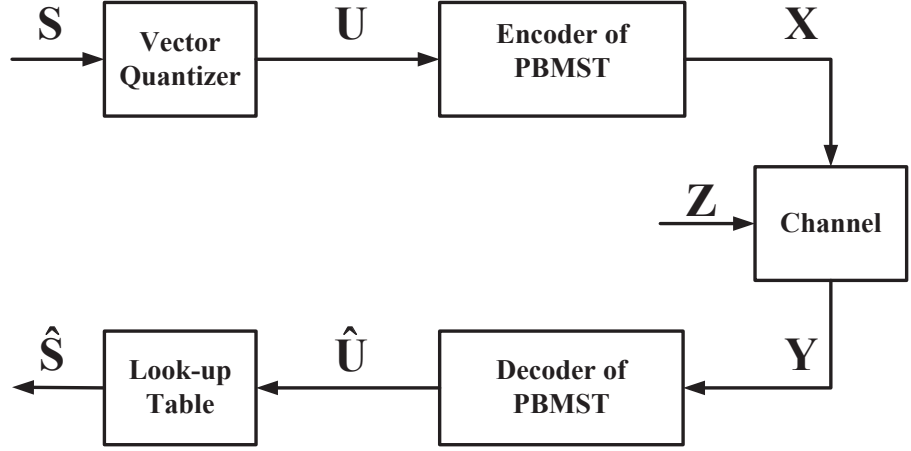}}
  \hspace{1in}
  \subfigure[Encoder of a partially block Markov superposition transmission system with memory $m$.]{
    \label{ProposedJSCC:subfig:b} 
    \includegraphics[width=12cm]{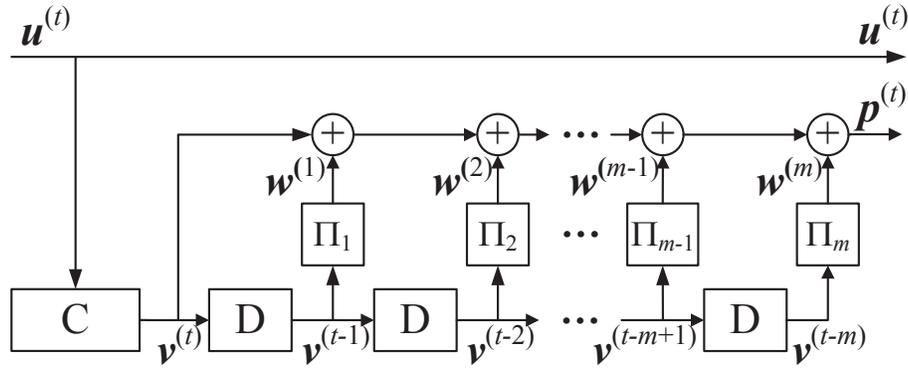}}
  \caption{Block diagram the proposed digital transmission schemes.}
  \label{ProposedJSCC:subfig} 
\end{figure}

Given a nested lattice code ${\rm G}$ and a systematic group code $\mathcal{C}[n,k]$ with $R_s/R_c\geq k/n$ defined on ${\rm G}$. We denote the codeword of $\mathcal{C}[n,k]$ corresponding to the information sequence $\mathbf{u}$ as $\mathbf{x}=(\mathbf{u},\mathbf{v})$, where $\mathbf{v}$ is the parity-check sequence. Let $\mathbf{s}^{(0)},\mathbf{s}^{(1)},\cdots,\mathbf{s}^{(L-1)}$ be $L$ blocks of source sequences, each of length-$k$. The proposed transmission scheme is described as follows, see Fig.~\ref{ProposedJSCC:subfig:a} and Fig.~\ref{ProposedJSCC:subfig:b} for reference.

\noindent\rule{16.6cm}{0.05cm}
{\bf Algorithm~1} Encoding of the Proposed JSCC Scheme

\noindent\rule{16.6cm}{0.02cm}
\begin{enumerate}
    \item{\bf{Initialization}:} For $t < 0$, set $\mathbf{v}^{(t)} = \mathbf{0} \in {\rm G}^{n-k}$.
    \item{\bf{Recursion}:} For $t = 0$, $1$, $\cdots$, $L-1$,
        \begin{itemize}
          \item \emph{Quantization:}~Quantize the $t$-th source block $\mathbf{s}^{(t)}$ based on ${\rm G}$, resulting the sequence $\mathbf{u}^{(t)}$.
          \item \emph{Encoding:}~Encode $\mathbf{u}^{(t)}$ with the encoding algorithm of the systematic group code $\mathcal{C}[n,k]$, the resulting parity-check sequence is denoted as $\mathbf{v}^{(t)} \in {\rm G}^{n-k}$;
          \item \emph{Interleaving:}~For $1\leq i \leq m$, interleave $\mathbf{v}^{(t-i)}$ by the $i$-th symbol interleaver
                $\Pi_{i}$ resulting in $\mathbf{w}^{(i)}$;
          \item \emph{Superposition:}~Compute $\mathbf{p}^{(t)} = \mathbf{v}^{(t)} + \sum_{1\leq i \leq m} \mathbf{w}^{(i)}$ and take $\mathbf{x}^{(t)}=(\mathbf{u}^{(t)},\mathbf{p}^{(t)})\in {\rm G}^n$ as the $t$-th transmitted block.
        \end{itemize}

    \item{\bf{Termination}:}
        For $t = L,L+1,\cdots,L+m-1$, set $\mathbf{s}^{(t)} = \mathbf{0}$ and compute $\mathbf{p}^{(t)}$ and $\mathbf{x}^{(t)}$ following Step.~2).
\end{enumerate}

\noindent\rule{16.6cm}{0.02cm}

Note that, when $m=0$, the parity-check sequences are transmitted without superposition. It can be seen that the proposed JSCC scheme avoids the use of an explicit entropy coding module. Note that a length $kL$ sequence is encoded into a length $n(L+m)$ sequence. Hence the bandwidth requirement is satisfied. For large $L$, the discrepancy between $\frac{kL}{n(L+m)}$ and $\frac{k}{n}$ is negligible. The addition operation of the group ${\rm G}$ is used in Step.~2). In the proposed scheme, only parity-check sequences are superimposed, hence termed partially block Markov superposition transmission~(PBMST). This is different from the original BMST scheme proposed in~\cite{Ma2015Block}. The motivation behind is explained as follows.
\begin{itemize}
  \item First, it was shown in~\cite{Shamai1998Systematic} that systematic JSCC schemes are optimal for a wide class of sources and channels.
  \item Second, in low SNR region, digital JSCC schemes may fail to decode and the resulting distortion would be large. If this is the case, the systematic part which is received directly from the channel can be taken as the decoding output. This ability to obtain the systematic part is critical for real-time applications and feedback-limited applications, such as multimedia streaming and deep-space communication~\cite{Bursalioglu2013Joint}.
\end{itemize}

\subsection{Decoding Algorithms of PBMST of Nested Lattice Codes}
\begin{figure}[!t]
\centering
\includegraphics[width=12cm]{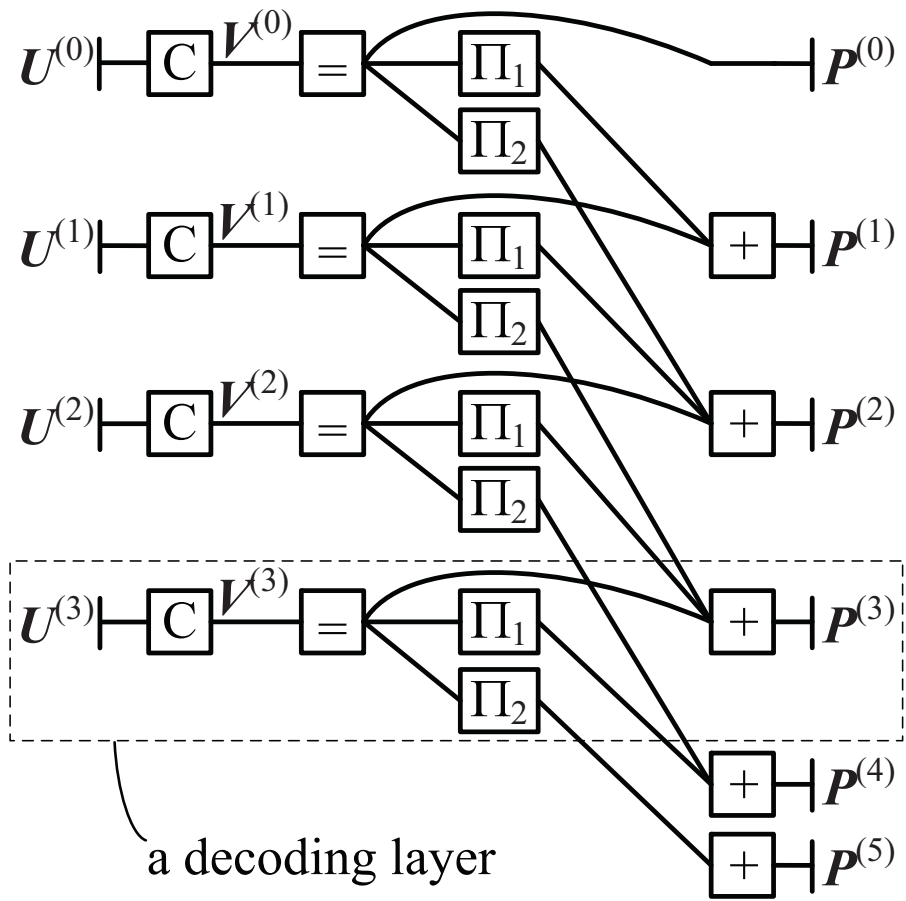}
\caption{The normal graphical representation of the proposed transmission scheme with $L=4$ and $m=2$.} \label{Decoder}
\end{figure}

The proposed transmission scheme can be decoded by a sliding-window decoding algorithm with a decoding delay delay $d$ over its normal graph, similar to the decoding algorithm of original BMST codes~\cite{Ma2015Block}. We show in Fig.~\ref{Decoder} a high-level normal graph of the proposed transmission scheme with $L=4$ and $m=2$. A decoding layer consists of nodes and edges in the dashed box of Fig.~\ref{Decoder}.

\noindent\rule{16.6cm}{0.05cm}
{\bf Algorithm~2} Iterative Sliding-Window Decoding Algorithm

\noindent\rule{16.6cm}{0.02cm}
\begin{itemize}
    \item{\bf{Global Initialization}:} Assume that $\mathbf{y}^{(t)}$, $0\le t \le L+m-1$, have been received. First, considering the channel transition probability and the initial probability in~(\ref{Initial}), compute the probability of the information sequence $P^{(|\rightarrow +)}_{U^{(t)}}(\mathbf{u}^{(t)})$. Second, consider only the channel transition probability, compute the {\em a posteriori} probability of the parity sequence $P^{(|\rightarrow +)}_{P^{(t)}}(\mathbf{p}^{(t)})$. All messages over the other edges within and connecting to the $t$-th layer~($0\le t\le d-1$) are initialized as uniformly distributed variables. Set a maximum iteration number $I_{max}>0$.

    \item{\bf{Recursion}:} For $t = 0$, $1$, $\cdots$, $L-1$,
        \begin{enumerate}
          \item {\bf\em Local initialization:}~If $t+d\le L+m-1$, compute $P^{(|\rightarrow +)}_{U^{(t)}}(\mathbf{u}^{(t)})$ and $P^{(|\rightarrow +)}_{P^{(t)}}(\mathbf{p}^{(t)})$ based on the received sequence $\mathbf{y}^{(t)}$. Initialize all the other edges within connecting to the $(t+d)$-th layer as uniformly distributed variables.
          \item {\bf\em Iteration:}~For $1\leq i \leq I_{max}$
          \begin{itemize}
          \item {\em Forward recursion:~}For $i=0,1,\cdots,\min(d,L+m-t-1)$, the $(t+i)$-th layer performs a message processing/passing algorithm scheduled as\\
              \fbox{+}$\rightarrow$\fbox{\footnotesize{$\Pi$}}$\rightarrow$\fbox{=}$\rightarrow$\fbox{\footnotesize{C}}$\rightarrow$\fbox{=}$\rightarrow$\fbox{\footnotesize{$\Pi$}}$\rightarrow$\fbox{+}.

          \item {\em Backward recursion:~}For $i=\min(d,L+m-t-1),\cdots,1,0$, the $(t+i)$-th layer performs a message processing/passing algorithm scheduled as\\
              \fbox{+}$\rightarrow$\fbox{\footnotesize{$\Pi$}}$\rightarrow$\fbox{=}$\rightarrow$\fbox{\footnotesize{C}}$\rightarrow$\fbox{=}$\rightarrow$\fbox{\footnotesize{$\Pi$}}$\rightarrow$\fbox{+}.
          \item {\em Hard decision:}~Make hard decisions on $\mathbf{u}^{(t)}$ resulting $\mathbf{\widehat{u}}^{(t)}$. If certain conditions are satisfied, output $\mathbf{\widehat{u}}^{(t)}$ and exit the iteration. Stopping criterion will be given later.
          \end{itemize}

          \item {\bf\em Cancelation:}~Remove the effect of $\mathbf{\widehat{v}}^{(t)}$ on all layers by updating the {\em a posteriori} probability as
              \begin{eqnarray*}
                {P}_{P_j^{(t+i)}}^{(|\rightarrow +)}(h)\leftarrow \sum_{g\in {\rm G}}{P}_{P_j^{(t+i)}}^{(|\rightarrow +)}(g){P}_{V_j^{t+i}}^{(\Pi_i\rightarrow +)}(h+\Pi_i(g)),
              \end{eqnarray*}
          for $h\in {\rm G}$, $j=0,1,\cdots,n-1$, and $i=1,2,\cdots,m$.
        \end{enumerate}
\end{itemize}

\noindent\rule{16.6cm}{0.02cm}

\noindent{\bf Remarks:~}Similar to the sliding-window decoding algorithm of spatial-coupled non-binary LDPC codes, the following entropy-based stopping criterion can be used in {\bf Algorithm~2}. Initially, we set a threshold $\epsilon>0$ and initialize the entropy rate $h_0(\mathbf{Y}^{(t)})=0$, where $\mathbf{Y}^{(t)}$ is the random vector corresponding to $\mathbf{y}^{(t)}$. For the $i$-th iteration, we estimate the entropy rate of $\mathbf{Y}^{(t)}$ by
\begin{eqnarray*}\label{Entropy}
    h_i(\mathbf{Y}^{(t)})=&-&\frac{1}{k}\sum_{j=0}^{k-1}\log\left(P_{Y_j^{(t)}}^{(C\rightarrow |)}(y_j^{(t)})\right)\\
    &-&\frac{1}{n-k}\sum_{j=0}^{n-k-1}\log\left(P_{Y_{k+j}^{(t)}}^{(+\rightarrow |)}(y_{k+j}^{(t)})\right)
\end{eqnarray*}
where $$P_{Y_j^{(t)}}^{(C\rightarrow |)}(y_j^{(t)})=\sum_{g\in {\rm G}}P_{U_j^{t}}^{(C\rightarrow |)}(g){\rm Pr}\{Y_j^{(t)}=y_j^{(t)}|U_j^{t}=g\}$$ for $0\le j\le k-1$ and $$P_{Y_{k+j}^{(t)}}^{(+\rightarrow |)}(y_{k+j}^{(t)})=\sum_{g\in {\rm G}}P_{P_j^{t}}^{(+\rightarrow |)}(g){\rm Pr}\{Y_{k+j}^{(t)}=y_{k+j}^{(t)}|P_{j}^{t}=g\}$$ for $1\le j\le n-k-1$. If $|h_i(\mathbf{Y}^{(t)})-h_{i-1}(\mathbf{Y}^{(t)})|<\epsilon$, exit the iteration.

\subsection{Complexity Analysis}
Let $q$ denote the cardinality of the nested lattice code ${\rm G}$. The decoding complexity of the sliding-window decoding algorithm is analyzed as follows. Let $Opt(A)$ denote the number of operations of the node $A$ in the normal graph. Each layer has $n-k$ parallel nodes $\fbox{=}$ of degree $m+2$, $n-k$ parallel nodes $\fbox{+}$ of degree $m+2$, and a node of type $\fbox{{\footnotesize C}}$. Hence, in each iteration, the total number of operations for each decoding layer is $(n-k) Opt(\fbox{+})+(n-k) Opt(\fbox{=})+Opt(\fbox{{\footnotesize C}})$. With trellis based implementation, we have $Opt(\fbox{=})=O(mq)$ and $Opt(\fbox{+})=O(mq^2)$. Decoding complexity of the node $\fbox{{\footnotesize C}}$ will be given in the next subsection.

\subsection{Design Considerations}
The proposed JSCC scheme is specified by the systematic group code $\mathcal{C}[n,k]$, the encoding memory $m$, and the decoding window size $d$. Performances of the proposed scheme are jointly determined by these factors. Typically, similar to the original BMST, we impose that the systematic group code $\mathcal{C}[n,k]$ has simple encoding algorithm and low-complexity maximum {\em a posteriori}~(MAP) decoding algorithm. We list in the following several possible candidates of $\mathcal{C}[n,k]$. Note that other systematic codes with efficient MAP decoding algorithm can also be used in the proposed scheme.
\begin{itemize}
  \item The $B$-fold Cartesian product of repetition group code, denoted as $\mathcal{C}_r[nB,B]\stackrel{\triangle}{=}[n,1]^B$, where $[n,1]$ represents the repetition code defined over ${\rm G}$. The encoder of $\mathcal{C}_r[nB,B]$ takes the sequence $\mathbf{u}$ of length-$B$ as input and outputs the sequence $(\mathbf{u},\mathbf{v})$ length-$nB$, where $\mathbf{v}=\underbrace{(\mathbf{u},\mathbf{u},\cdots,\mathbf{u})}_{n-1}$. The MAP decoding algorithm of $\mathcal{C}[nB,B]$ can be implemented on a $q$-state trellis with decoding complexity $O(nqB)$.
  \item The $B$-fold Cartesian product of single parity-check group code, denoted as $\mathcal{C}_s[nB,(n-1)B]\stackrel{\triangle}{=}[n,n-1]^B$, where $[n,n-1]$ represents the single parity-check code defined over $G$. The encoder of $\mathcal{C}_s[nB,(n-1)B]$ takes the sequence $\mathbf{u}=(\mathbf{u}_0,\mathbf{u}_1,\cdots,\mathbf{u}_{n-2})$ of length-$(n-1)B$ as input and outputs the sequence $(\mathbf{u},\mathbf{v})$ of length-$nB$, where $\mathbf{v}=\sum_{i=0}^{n-2}\mathbf{u}_i$. The MAP decoding algorithm of $\mathcal{C}[nB,(n-1)B]$ can be implemented on a $q$-state trellis with decoding complexity $O(nq^2B)$.
  \item The group code $\mathcal{C}[n(B_1+B_2),B_1(n-1)+B_2]$ formed by time-sharing between $\mathcal{C}_s[nB_1,(n-1)B_1]$ and $\mathcal{C}_r[nB_2,B_2]$.
\end{itemize}

Given the group code $\mathcal{C}[n,k]$, the encoding memory $m$ along with the decoding window size $d$ have impact not only on the performances but also on the complexities. For different systematic group codes, the required encoding memories to achieve the optimal performance are different. From our numerical results, we find that $m\geq10$ is sufficient. Considering the window size, larger window size $d$ typically corresponds to better performance. In practice, acceptable performances are obtained by setting $d=2m\sim 3m$.

\section{PBMST with One-Dimensional Nested Lattice Codes}\label{SecIV}

\begin{table*}
\caption{Optimized quantizer based on $\mathbb{Z}/3\mathbb{Z}$ for memoryless Gaussian source~($\ell=1,\alpha=1.22$).}\label{one-d-3} \centering
\begin{threeparttable}
\begin{tabular}{l||l|l|l}
\hline Intervals & $(-\infty , -0.61] $& $(-0.61 , 0.61]$ & $(0.61 , +\infty)$ \\
 \hline Representative points & -1.22 & 0 & 1.22 \\
\hline Probability & 0.2709 & 0.4582 & 0.2709 \\
\hline
\end{tabular}
\begin{tablenotes}[para]
Entropy : ${\rm H}=1.53$~bits/symbol~~~~~~~~~~       Distortion : $D=0.190$.
\end{tablenotes}
\end{threeparttable}
\end{table*}

\begin{table*}
\caption{Optimized quantizer based on $\mathbb{Z}/5\mathbb{Z}$ for memoryless Gaussian source~($\ell=1,\alpha=0.83$).}\label{one-d-5} \centering
\begin{threeparttable}
\begin{tabular}{l||l|l|l|l|l}
\hline Intervals & $(-\infty , -1.24] $& $(-1.24 , -0.41]$ & $(-0.41 , 0.41)$  & $(0.41 , 1.24)$& $(1.24 , +\infty)$\\
 \hline Representative points & -1.66 & -0.83 & 0 & 0.83 &  1.66\\
\hline Probability & 0.1075 & 0.2334 & 0.3182 & 0.2334 &   0.1075\\
\hline
\end{tabular}
\begin{tablenotes}[para]
Entropy : ${\rm H}=2.19$~bits/symbol~~~~~~~~~~       Distortion : $D=0.082$.
\end{tablenotes}
\end{threeparttable}
\end{table*}

\begin{figure}[!t]
\centering
\includegraphics[width=12cm]{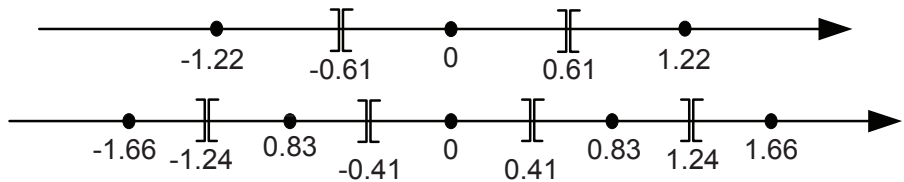}
\caption{The representative points and their corresponding Voronoi regions of the optimized quantizer based on $\mathbb{Z}/3\mathbb{Z}$ and $\mathbb{Z}/5\mathbb{Z}$.} \label{one_dimensional_quantizer}
\end{figure}

\begin{figure}
\centering
\includegraphics[width=12cm]{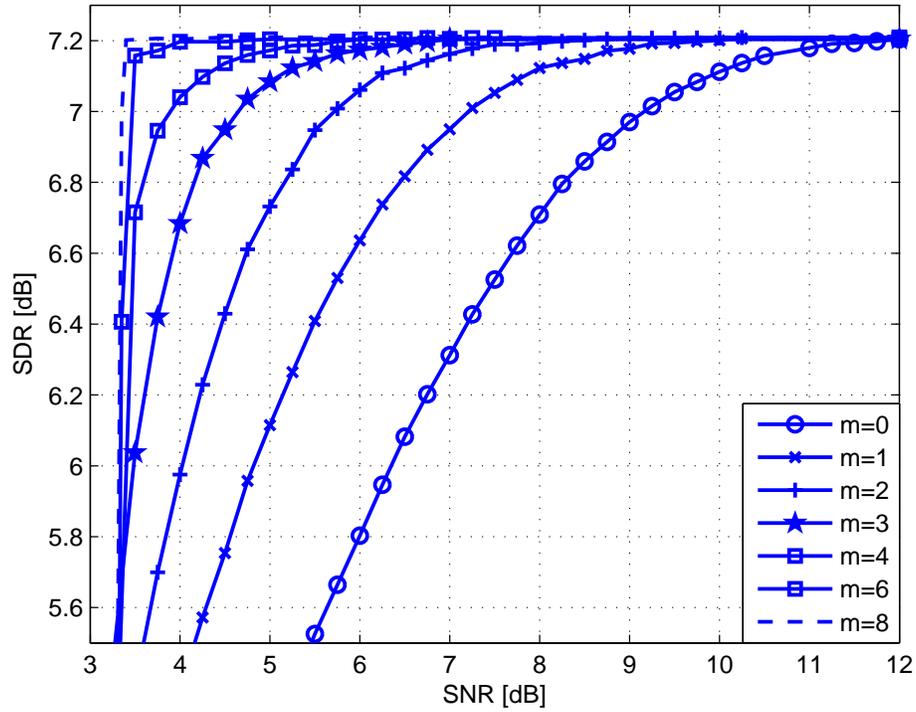}
\caption{The performances of the proposed JSCC scheme:~$\mathbb{Z}/3\mathbb{Z}$ and memoryless Gaussian source.} \label{Z3PvsD}
\end{figure}

\begin{figure}
\centering
\includegraphics[width=12cm]{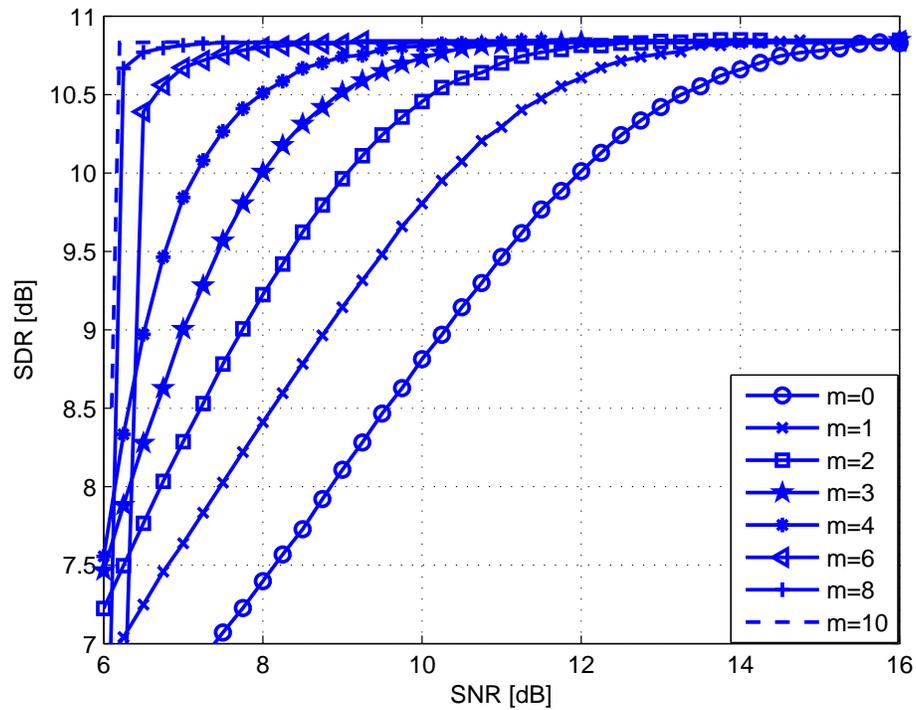}
\caption{The performances of the proposed JSCC scheme:~$\mathbb{Z}/5\mathbb{Z}$ and memoryless Gaussian source.} \label{Z5PvsD}
\end{figure}

For $\ell=1$, we take the nested lattice codes $\mathbb{Z}/3\mathbb{Z}$ and $\mathbb{Z}/5\mathbb{Z}$ for illustrations. For each nested lattice code, the optimized lattice quantizer is obtained by searching the scaling factor $\alpha$ to minimize the distortion. The parameters of the optimized lattice quantizers are shown in Table~\ref{one-d-3} and Table~\ref{one-d-5}. As expected the distortion of the lattice quantizer based on $\mathbb{Z}/5\mathbb{Z}$ is lower than the distortion of the quantizer based on $\mathbb{Z}/3\mathbb{Z}$. Graphical illustrations of the optimized lattice quantizers based on $\mathbb{Z}/3\mathbb{Z}$ and $\mathbb{Z}/5\mathbb{Z}$ are shown in Fig.~\ref{one_dimensional_quantizer}. We assume that $R_s/R_c=1/2$. The systematic group code is selected as the $1000$-fold Cartesian product repetition code $C_r[2000,1000]$. The simulation results for different encoding memories are shown in Fig.~\ref{Z3PvsD} and Fig.~\ref{Z5PvsD}, where the distortion is measured by the normalized squared errors~(average over frames) between the output of the source and the output from the decoder, as defined in~(\ref{Distortion}). The simulation results are shown in terms of signal-to-distortion ratio~(SDR), defined as
\begin{equation}\label{SDR}
    {\rm SDR}=\frac{1}{D}.
\end{equation}

In the following, performances of the proposed JSCC scheme is analyzed and compared. For digital transmission system with quanziter based on $\mathbb{Z}/3\mathbb{Z}$, the source encoder takes $R_s$ {\em symbols/T} as input and delivers at least $R_sH = 1.53R_s$ {\em bits/T} as output. Hence the channel coding rate is at least $0.765$~{\em bits/channel-use}, which implies that the required SNR for reliable digital transmission must be greater than $2.76$~dB. This analysis shows that the proposed JSCC scheme performs $0.59$~dB away from the theoretical limit, but with finite length and simple implementation.

For digital transmission system with quanziter based on $\mathbb{Z}/5\mathbb{Z}$, the source encoder takes $R_s$ {\em symbols/T} as input and delivers at least $R_sH = 2.19R_s$ {\em bits/T} as output. Hence the channel coding rate is at least $1.095$~{\em bits/channel-use}, which implies that the required SNR for reliable digital transmission must be greater than $5.52$~dB. This analysis shows that the proposed JSCC scheme performs $0.68$~dB away from the theoretical limit.

We also compare the proposed scheme with the optimal performance theoretically attainable~(OPTA)~\cite{Akyol2014On}. For $D=0.19$, according to the rate-distortion function of Gaussian sources~\cite{Cover91}, at least $1.198$ bits are required to represent each source symbol. Hence, the required SNR for reliable transmission, without constraints on the quantizer, the source encoder, and the channel encoder, is 1.20~dB. It can be seen that the proposed transmission scheme based on $\mathbb{Z}/3\mathbb{Z}$ performs 2.15~dB away from the unconstrained theoretical limit. Similar analysis shows that the proposed transmission scheme based on $\mathbb{Z}/5\mathbb{Z}$ performs 2.0~dB away from the unconstrained theoretical limit. From the above analysis, we conclude that the proposed scheme performs well with respect to both the constrained and the unconstrained limits.


\section{PBMST with Two-Dimensional Nested Lattice Codes}\label{SecV}
\subsection{Optimization of Two-dimensional Nested Lattice Codes}
For two-dimensional nested lattice code, two consecutive symbols from the source are quantized jointly, resulting in a point in the nested lattice code. The following algorithm can be implemented to optimize the parameters of the lattice quantizer to lower down the distortion.

\noindent\rule{16.6cm}{0.05cm}
{\bf Algorithm~3} An Iterative Optimization Algorithm of Two-dimensional Nested Lattice Code

\noindent\rule{16.6cm}{0.02cm}
\begin{enumerate}
    \item {\bf Initialization:}~Set $\alpha = 1$ and $\mathbf{v} = (0,0)$. Select the parameter $\delta$.

    \item {\bf Iteration:}
   \begin{enumerate}
     \item {\em Scaling:}~scale the lattice to find $\alpha$ which results in the least distortion $D$ with shifting vector equal to $\mathbf{v}$. This is an one-dimensional optimization problem and can be solved by existing optimization algorithms.
     \item {\em Shifting:}~choose the four shifting vectors $\mathbf{v}_1=(\sqrt2/2,\sqrt2/2)$, $\mathbf{v}_2=(-\sqrt2/2,\sqrt2/2)$, $\mathbf{v}_3=(\sqrt2/2,-\sqrt2/2)$, $\mathbf{v}_4=(-\sqrt2/2,-\sqrt2/2)$ and the step size $\delta$. Compute the distortion, denoted by $D_i$, of the nested lattice code $\alpha (A_2+\delta \mathbf{v}_i)/3\alpha A_2$. Find the minimal distortion $D'$ as
         \begin{equation}
            D'=\min_{1\leq i \leq 4}{D_i}.
         \end{equation}
         The  shifting vector associated with $D'$ is denoted by $\mathbf{v}'$. If $D'<D$, set $\mathbf{v}=\delta \mathbf{v}'$, go to Step 2~(a).
     \item {\em Bi-section:}~set $\delta=\delta/2$, if $\delta< 0.001$, exit the iteration; else go to Step 2~(b).
   \end{enumerate}
\end{enumerate}
\noindent\rule{16.6cm}{0.02cm}
\noindent{\bf Remark:}~Note that in { Step 2~b)} only one direction is chosen for shifting in each iteration. In our simulations, the parameter $\delta$ is set to be $1.0$.

\subsection{Nested Lattice Code Based on Hexagonal Lattice}
\begin{figure}[!t]
\centering
\includegraphics[width=12cm]{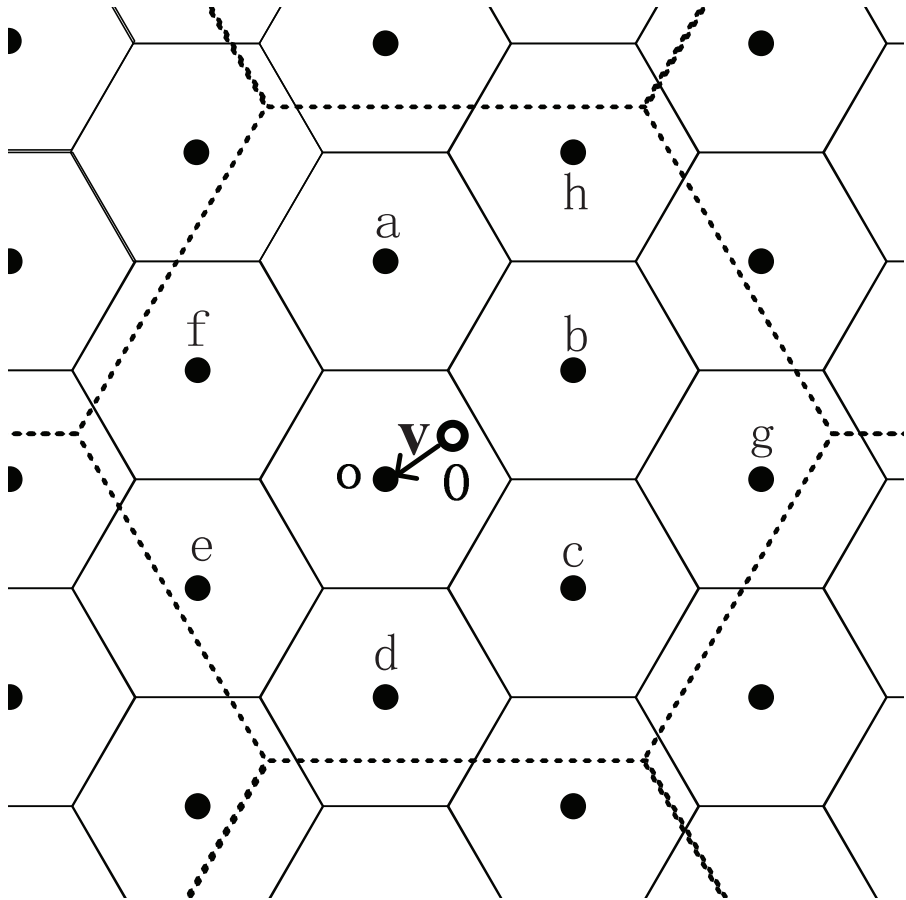}
\caption{The nested lattice code based on $A_2/3A_2$:~$A_2$ is the hexagonal lattice, $\mathbf{v}$ is the shifting vector and the shaping region is the doted hexagon.} \label{lattice}
\end{figure}

For $\ell=2$, a nested lattice code based on the two-dimensional hexagonal lattice $A_2$ is selected for simulation. The sublattice chosen for shaping is $3A_2$. The lattice points in $A_2$ are integer combinations of the base vectors $\mathbf{g}_1 = (1/2, \sqrt{3}/6)$, $\mathbf{g}_2 = (1/2, -\sqrt{3}/6)$. That is $A_2=\{\alpha_1\mathbf{g}_1+\alpha_2\mathbf{g}_2:~\alpha_1\in \mathbb{Z}\textrm{~and~}\alpha_2\in \mathbb{Z}\}$. The sublattice $3A_2$ is generated by basis vectors $\mathbf{g}_3= (3/2,\sqrt3/2), \mathbf{g}_4=(3/2,-\sqrt3/2)$. The nested lattice code is shown in Fig.~\ref{lattice}. Note that only the points in the doted hexagon are chosen as codewords of the nested lattice code. According to the labels in Fig~\ref{lattice}, we can define a finite abelian group ${\rm G}\stackrel{\Delta}{=}\{{\rm {\bf o},a,b,c,d,e,f,g,h}\}$ with the operation rule defined in Table~\ref{add_list}. The identity element is \textbf{o}. Actually, the group ${\rm G}$ is the quotient group defined by $A_2$ and its subgroup $3A_2$, that is, ${\rm G}=A_2/3A_2$.

\begin{table}
\caption{The addition table of the quotient group $A_2/3A_2$, where \textbf{o} is the identity element.}\label{add_list} \centering
\begin{tabular}{l||l|l|l|l|l|l|l|l|l}
\hline   $+$  &a &b &c &d &e &f &g &h &o \\
\hline
 \hline  a   &e &h &b &\textbf{o} &f &g &e &c &a \\
 \hline  b   &h &e &g &c &\textbf{o} &a &d &f &b \\
 \hline  c   &b &g &f &h &d &\textbf{o} &a &e &c \\
 \hline  d   &\textbf{o} &c &h &a &g &e &f &b &d \\
 \hline  e   &f &\textbf{o} &d &g &b &h &c &a &e \\
 \hline  f   &g &a &\textbf{o} &e &h &c &b &d &f \\
 \hline  g   &e &d &a &f &c &b &h &\textbf{o} &g \\
 \hline  h   &c &f &e &b &a &d &\textbf{o} &g &h \\
 \hline  \textbf{o}   &a &b &c &d &e &f &g &h &\textbf{o} \\
\hline
\end{tabular}
\end{table}

\subsubsection{Memoryless Gaussian Sources}
\begin{figure}[!t]
\centering
\includegraphics[width=12cm]{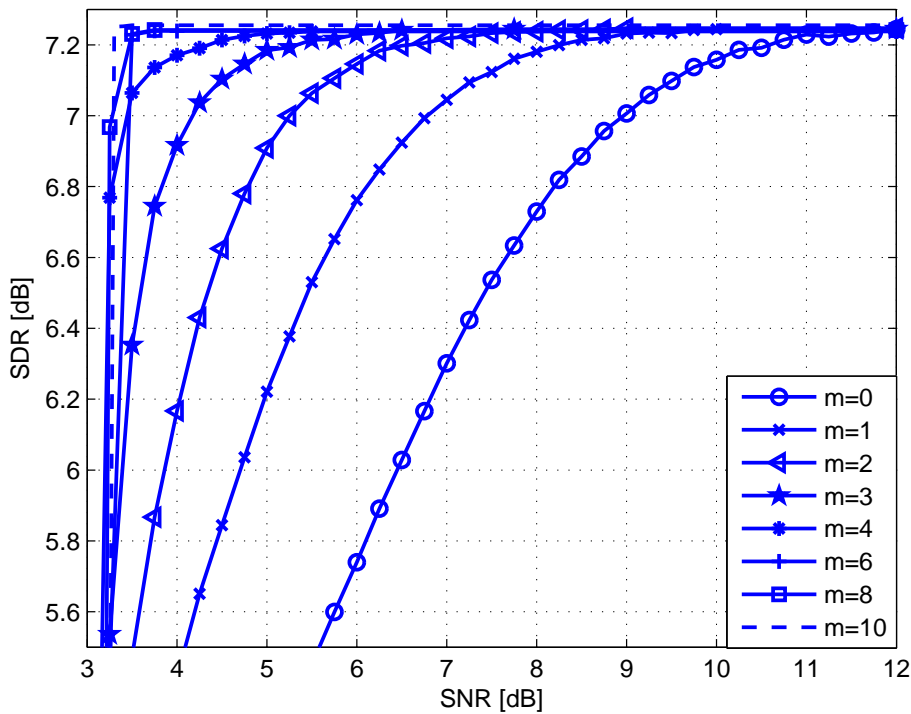}
\caption{The performances of the proposed JSCC scheme:~$A_2/3A_2$ and memoryless Gaussian source.} \label{A2MemorylessPvsD}
\end{figure}

We have simulated the proposed scheme based on two-dimensional nested lattice code for memoryless Gaussian source. {\bf Algorithm 3} is implemented to optimize the inflation factor $\alpha$ and the shifting vector $\mathbf{v}$. The results are shown in Table~\ref{twoDMemoryless}. Also shown in Table~\ref{twoDMemoryless} are the coordinates of the representative points. It can be seen that two-dimensional quantizer reveals lower distortion than one-dimensional quantizer. We assume that $R_S/R_C = 1/2$. The systematic group code is selected as the $1000$-fold Cartesian product repetition code $C_r[2000,1000]$. The simulation results for different encoding memories are shown in Fig.~\ref{A2MemorylessPvsD}.

For digital transmission system with quanziter based on ${A}_2/3{A_2}$, the source encoder takes $R_s$ {\em symbols/T} as input and delivers at least $\frac{R_s}{2}H = 1.525R_s$ {\em bits/T} as output. Hence the channel coding rate is at least $0.763$~{\em bits/channel-use}, which implies that the required SNR of digital transmission scheme must be greater than $2.73$~dB. This analysis shows that the proposed JSCC scheme performs $0.5$~dB away from the theoretical limit, but with finite length and simple implementation.

It can be seen from Fig.~\ref{A2MemorylessPvsD} that the proposed scheme based on two-dimensional lattice performs about 0.15~dB better than that based on one-dimensional lattice. Also note that distortion of two-dimensional lattice based scheme is slightly lower than that of one-dimensional lattice based scheme. This is consistent with the results in Table~\ref{one-d-3} and Table~\ref{twoDMemoryless}. However, the two-dimensional lattice based scheme requires more computation than the scheme based on one-dimensional lattice.

\begin{table}
\caption{Optimized vector quantizer based on two-dimensional lattice for memoryless Gaussian source~($\ell=2,\alpha=2.26,\mathbf{v}=(-0.151,-0.088)$).}\label{twoDMemoryless} \centering
\begin{tabular}{l|l|l}
\hline                symbol &  coordinates  &  probability \\
\hline                a &$(-0.3413  ,     1.1059)$  &         0.1369  \\
 \hline               b &$(0.7887   ,     0.4535)$  &      0.1529   \\
 \hline                c &$(0.7887  ,      -0.8513)$ &      0.1361   \\
 \hline                d &$(-0.3413 ,      -1.5037)$ &      0.0940   \\
 \hline                e &$(-1.4713  ,      -0.8513)$ &     0.0716   \\
 \hline                f &$(-1.4713 ,      0.4535)$ &   0.0942     \\
 \hline                g &$(1.9187  ,     -0.1989)$ &     0.0597    \\
 \hline                h &$(0.7887  ,      1.7583)$ &    0.0595     \\
  \hline                o &$(-0.3413 ,      -0.1989)$           &     0.1951     \\
\hline
\end{tabular}
\begin{tablenotes}[para]
Entropy : ${\rm H}=3.05$~bits/two-dimension

Distortion : $D=0.188$.
\end{tablenotes}
\end{table}

\subsubsection{Correlated Gaussian Sources}
\begin{figure}[!t]
\centering
\includegraphics[width=12cm]{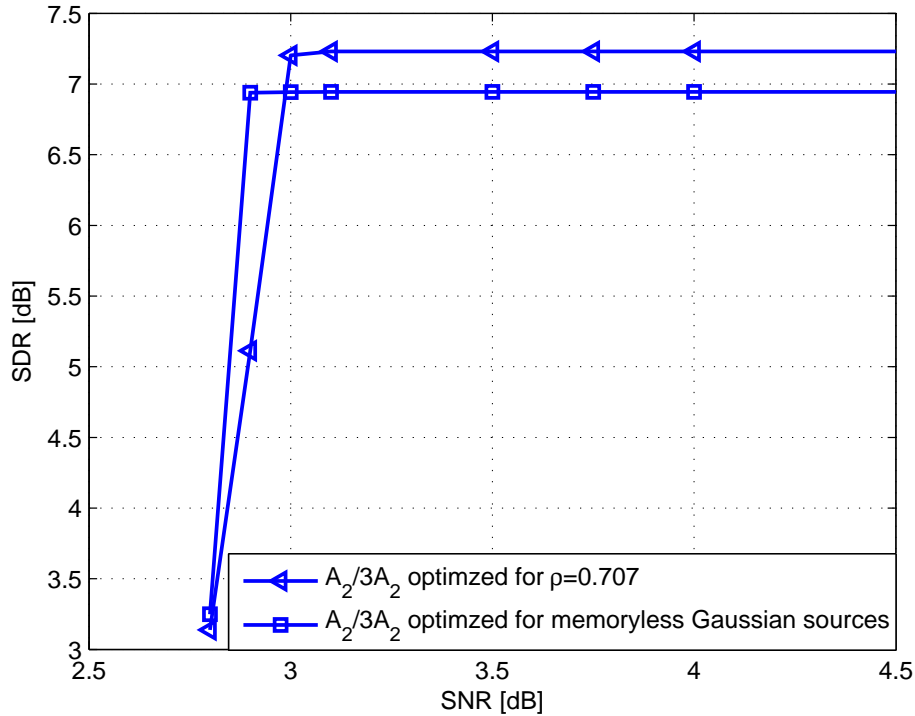}
\caption{The performances of the proposed JSCC scheme:~two-dimensional lattice quantizers, $m=8$, and correlated Gaussian source.} \label{correlated}
\end{figure}

Consider a correlated discrete-time Gaussian source $\mathbf{S}=(S_0, S_1, S_2,...)$, where $S_i$ is a Gaussian random variable with zero
mean and variance $\sigma^2 = 1$, defined as
\begin{equation}\label{corr}
    S_{2k+1}=\rho S_{2k} + \sqrt{1-\rho^2}Z_{2k},
\end{equation}
where $Z_{2k}$ is a Gaussian random variable with zero mean and variance $\sigma^2=1$ and $\rho\in[0,1]$ is the {\em correlation factor}. Note that $\rho=1$ corresponds to completely correlated Gaussian source, while $\rho=0$ corresponds to memoryless Gaussian source. For different correlation factors, {\bf Algorithm 3} is executed to find optimized lattice quantizers. The optimization results are shown in Table~\ref{correlated707} for $0.707$. We assume that $R_S/R_C = 1/2$. The systematic group code is selected as the $1000$-fold Cartesian product repetition code $C_r[2000,1000]$. The simulation results are shown in Fig.~\ref{correlated} for $m=8$.


For conventional digital transmission system with quanziter based on ${A}_2/3{A_2}$, the source encoder takes $R_s$ {\em symbols/T} as input and delivers at least $\frac{R_s}{2}H = 1.375R_s$ {\em bits/T} as output. Hence the channel coding rate is at least $0.6875$~{\em bits/channel-use}, which implies that the required SNR of digital transmission scheme must be greater than $2.02$~dB. This analysis shows that the proposed JSCC scheme performs $0.88$~dB away from the theoretical limit, but with finite length and simple implementation.

For comparison, we have also simulated the proposed JSCC scheme with the lattice quantizer in Table~\ref{twoDMemoryless}~(optimized for memoryless Gaussian sources) for the transmission of correlated Gaussian sources. Simulation results are shown in Fig.~\ref{correlated} with $m=8$. It can be seen from Fig.~\ref{correlated} that, for correlated Gaussian sources, the proposed schemes based on optimized nested lattice codes reveal lower distortions than the scheme based on nested lattice code designed for memoryless Gaussian sources. This shows that the optimization of nested lattice codes is critical to the proposed JSCC scheme.

%

\begin{table}
\caption{Optimized vector quantizer based on two-dimensional lattice for a correlated Gaussian source: $\rho=\sqrt2/2$. ($m=2,\alpha=2.27,\mathbf{v}=(-0.0375,-0.225)$).}\label{correlated707} \centering
\begin{tabular}{l|l|l}
\hline                symbol &  coordinates  &  probability \\
\hline                a &$(-0.0855  ,     0.8034)$  &         0.1660  \\
 \hline               b &$(1.0545   ,     0.1452)$  &      0.1454   \\
 \hline                c &$(1.0545  ,      -1.1712)$ &      0.0135   \\
 \hline                d &$(-0.0855 ,      -1.8294)$ &      0.0310   \\
 \hline                e &$(-1.2255  ,      -1.171)$ &     0.1696   \\
 \hline                f &$(-1.2255 ,      0.1452)$ &   0.0817     \\
 \hline                g &$(2.194  ,     -0.5130)$ &     0.0047    \\
 \hline                h &$(1.0545  ,      1.4614)$ &    0.1559     \\
  \hline                o &$(-0.0855 ,      -0.5130)$           &     0.2322     \\
\hline
\end{tabular}
\begin{tablenotes}[para]
Entropy : ${\rm H}=2.75$~bits/two-dimension

Distortion : $D=0.189$.
\end{tablenotes}
\end{table}

%
%

\section{Conclusion}\label{SecVI}
We studied in this paper the digital transmission of Gaussian sources, memoryless or correlated, through AWGN channels in bandwidth expansion regime. We proposed a new JSCC scheme based on the partially block Markov superposition transmission of nested lattice codes. Advantages of the proposed scheme include:~1)~It avoids the explicit use of entropy coding module, hence mitigates the error propagation phenomenon of conventional pure digital scheme. 2)~Its performs well for different nested lattice codes and different Gaussian sources. The proposed scheme provides an attractive candidate for digital transmission of Gaussian sources through AWGN channels, either in purely digital systems or in HDA systems.


%

\section*{Acknowledgment}
The authors would like to thank Mr.~Zheng Yang for helpful discussion.

\appendices



%
%

\ifCLASSOPTIONcaptionsoff

\newpage
\fi



%


\bibliographystyle{IEEEtran}
\bibliography{IEEEabrv,tzzt}




%

%
%
%




\end{document}